\documentclass[aps,pra,twocolumn,superscriptaddress]{revtex4-1}

\usepackage{graphicx}
\usepackage{dcolumn}
\usepackage{bm}
\usepackage{amsmath}
\usepackage{amsmath,bm}
\usepackage{bbold}

\begin{document}

\preprint{APS/123-QED}

\title{Quaternion algebra for Stokes-Mueller formalism}

\author{Ertan Kuntman}
\affiliation{Departament de F\'isica Aplicada, Institute of Nanoscience and Nanotechnology (IN2UB), Feman Group, Universitat de Barcelona, C/ Mart\'{\i} i Franqu\`es 1, Barcelona 08030, Spain.}
\author{M. Ali Kuntman}
\email{makuntman@gmail.com}
\affiliation{Independent researcher, Ankara, Turkey\\
}
\author{Oriol Arteaga}
\affiliation{Departament de F\'isica Aplicada, Institute of Nanoscience and Nanotechnology (IN2UB), Feman Group, Universitat de Barcelona, C/ Mart\'{\i} i Franqu\`es 1, Barcelona 08030, Spain.}

\date{\today}

\begin{abstract}
It is shown that the Stokes-Mueller formalism can be reformulated in terms of quaternions, and the quaternion approach is more suitable for the formalism of Mueller-Jones states that we have recently described. In terms of quaternions it can be shown that the vector and matrix states and the Jones matrix associated to nondepolarizing optical systems  are different representations isomorphic to the same quaternion state, and this quaternion state turns out to be the rotator of the Stokes quaternion. It is also shown that the coherent linear combination of nondepolarizing optical media states and depolarization phenomena can be reformulated in terms of quaternion states.
\end{abstract}

\pacs{Valid PACS appear here}
\maketitle


\section{Introduction}
In our previous works \cite{KKPA,KKO},  we have introduced
two different representations of Mueller-Jones states: the covariance vector $|h\rangle$ and the $\mathbf{Z}$ matrix. Covariance vector, $|h\rangle$,
can be defined in terms of the covariance matrix $\mathbf{H}$ associated to the Mueller matrix $\mathbf{M}$:

\begin{equation}
\mathbf{H}= \frac{1}{4}\sum_{i,j=0}^3M_{ij}{\Pi}_{ij},
\end{equation}
where $M_{ij} (i,j = 0,1,2,3)$ are the elements of the Mueller matrix
and ${\Pi}_{ij}=\mathbf{A}(\bm{{\sigma}_{i}}\otimes\bm{{\sigma}_{j}}^*)\mathbf{A^{-1}}$.

{\footnotesize
\begin{equation}
\mathbf{A}=\begin{pmatrix}
1&0&0&1\\
1&0&0&-1\\
0&1&1&0\\
0&i&-i&0\end{pmatrix},
\quad\mathbf{A}^{-1}=\frac{1}{2}\mathbf{A}^{\dagger}=\frac{1}{2}\begin{pmatrix}
1&1&0&0\\
0&0&1&-i\\
0&0&1&i\\
1&-1&0&0
\end{pmatrix}.
\end{equation}}
The superscript $^{\dagger}$ indicates the complex conjugate and transpose, the superscript $^*$ indicates complex conjugate, $\otimes$ is the Kronecker product and $\bm{\sigma_{i}}$ are the Pauli matrices with the $2\times2$ identity in the following order:
\begin{align}
\bm{{\sigma}_0}=\begin{pmatrix}
1&0\\
0&1
\end{pmatrix},\quad\bm{{\sigma}_1}=\begin{pmatrix}
1&0\\
0&-1
\end{pmatrix},\\\bm{{\sigma}_2}=\begin{pmatrix}
0&1\\
1&0
\end{pmatrix},\quad\bm{{\sigma}_3}=\begin{pmatrix}
0&-i\\
i&0
\end{pmatrix},
\end{align}

If and only if the Mueller matrix of the system is nondepolarizing, the associated covariance matrix $\mathbf{H}$ will be of rank 1. In this case it is always possible to define a covariance vector $|h\rangle$ such that\cite{Gil1, Gil2, Aelio}:

\begin{equation}
\mathbf{H}=| h\rangle\langle h|,
\end{equation}
where $|h\rangle$ is the eigenvector of $\mathbf{H}$ corresponding to the single nonzero eigenvalue.

In the preferred basis ${\Pi}_{ij}$ the dimensionless components of $|h\rangle$ can be parametrized as $\tau$, $\alpha$, $\beta$, $\gamma$:
\begin{equation}
|h\rangle=\begin{pmatrix}
 \tau\\\alpha\\\beta\\\gamma
\end{pmatrix},
\end{equation}
where $\alpha$, $\beta$ and $\gamma$ are generally complex numbers, while $\tau$ can always be chosen as real and positive if we discard the global phase. 

On the other hand, in our previous works there was another object, $\mathbf{Z}$, that serves as a Mueller-Jones state in matrix form:

\begin{equation}
\mathbf{Z}=\begin{pmatrix}
\tau&\alpha&\beta&\gamma\\
\alpha&\tau&-i\gamma&i\beta\\
\beta&i\gamma&\tau&-i\alpha\\
\gamma&-i\beta&i\alpha&\tau
\end{pmatrix}.
\end{equation}

By direct matrix multiplication it can be shown that the Mueller matrix of any nondepolarizing optical media can be written as:
\begin{equation}\label{commute}
\mathbf{M}= \mathbf{Z}\mathbf{Z^*}=\mathbf{Z^*}\mathbf{Z}.
\end{equation}

The Mueller matrix transforms the stokes vector $|s\rangle=(s_0,s_1,s_2,s_3)^T$ into $|s'\rangle$:

\begin{equation}
|s'\rangle=\mathbf{M}|s\rangle
\end{equation}.

It can be shown that the $\mathbf{Z}$ matrix transforms the Stokes matrix according to the following scheme:

\begin{equation}\label{ZSZ}
\mathbf{S'}= \mathbf{Z}\mathbf{S}\mathbf{Z^\dagger}
\end{equation}
where the Stokes matrix, $\mathbf{S}$, is defined as
\begin{equation}\label{squat}
\mathbf{S}=\begin{pmatrix}
s_0 &s_1 &s_2 &s_3\\
s_1 &s_0 &-is_3 &is_2\\
s_2&is_3&s_0&-is_1\\
s_3&-is_2&is_1&s_0
\end{pmatrix}
\end{equation}

The $\mathbf{Z}$ matrices are also suitable to describe the product state of the combined system associated to a serial combination of the optical media: 

\begin{equation}\label{mmmu}
\mathbf{Z}= \mathbf{Z}_{N}\cdot\mathbf{Z}_{N-1}\cdots\mathbf{Z}_{2}\cdot\mathbf{Z}_{1}.
\end{equation}

But similar algebra is not possible with $|h\rangle$ vectors. Therefore, $|h\rangle$ vectors and $\mathbf{Z}$ matrices appear as different entities in their present form. 

In this work it will be shown that $\mathbf{Z}$ matrices and $|h\rangle$  vectors are actually two different representations of the same quantity which are isomorphic to the $\bm{h}$ quaternion.

\section{quaternion state}

First, we observe that the $\mathbf{Z}$ matrix can be written as a linear combination of four matrices:

\begin{equation}\label{zmat}
\mathbf{Z}=\tau \mathbb{1}+i\alpha I+i\beta J+i\gamma K,
\end{equation}
where

\begin{equation*}
\mathbb{1}=
\begin{pmatrix}
1&0&0&0\\
0&1&0&0\\
0&0&1&0\\
0&0&0&1
\end{pmatrix},\qquad
I=\begin{pmatrix}
0&-i&0&0\\
-i&0&0&0\\
0&0&0&-1\\
0&0&1&0
\end{pmatrix}\qquad
\end{equation*}

\begin{equation}
J=
\begin{pmatrix}
0&0&-i&0\\
0&0&0&1\\
-i&0&0&0\\
0&-1&0&0
\end{pmatrix},\qquad
K=\begin{pmatrix}
0&0&0&-i\\
0&0&-1&0\\
0&1&0&0\\
-i&0&0&0
\end{pmatrix}\qquad
\end{equation}

These basis matrices have the following properties:

\begin{equation*}
I^2= J^2=K^2=IJK=-\mathbb{1}\end{equation*}
\begin{equation}
IJ=-JI=K,\quad JK=-KJ=I, \quad KI=-IK=J\end{equation}

It is clear that these matrices  are isomorphic to the quaternion basis defined by Hamilton \cite{Hamilton}:

\begin{equation*}
\bm{i}^2=\bm{j}^2=\bm{k}^2=\bm{i}\bm{j}\bm{k}=-1
\end{equation*}
\begin{equation}
\bm{i}\bm{j}=-\bm{j}\bm{i}=\bm{k},\quad\bm{j}\bm{k}=-\bm{k}\bm{j}=\bm{i},\quad\bm{k}\bm{i}=-\bm{i}\bm{k}=\bm{j}
\end{equation}

 Therefore, the $\mathbf{Z}$ matrix of \eqref{zmat} is isomorphic to the $\bm{h}$ quaternion:

\begin{equation}
\bm{h}=\tau1+i\alpha\bm{i}+i\beta\bm{j}+i\gamma\bm{k},
\end{equation}
which is directly related to the covariance vector $|h\rangle$.

It is worth noting that the Jones matrix is also isomorphic to the quaternion $\bm{h}$. In order to show this we write the Jones matrix in terms of Pauli matrices \cite{KKO}:
\begin{equation}
\mathbf{J}= \tau\bm{\sigma}_0+\alpha\bm{\sigma}_1+\beta\bm{\sigma}_2+\gamma\bm{\sigma}_3.
\end{equation}
which can be written as
\begin{equation}
\mathbf{J}= \tau\bm{\sigma}_0+i\alpha(-i\bm{\sigma}_1)+i\beta(-i\bm{\sigma}_2)+i\gamma(-i\bm{\sigma}_3).
\end{equation}
The 2$\times$2 matrices $\bm{\sigma}_0$, $-i\bm{\sigma}_1$, $-i\bm{\sigma}_2$ and $-i\bm{\sigma}_3$ are, respectively, isomorphic to the quaternion basis, 1, $\bm{i},\bm{j}$ and $\bm{k}$. Therefore, we can write the associated Jones quaternion as follows:
\begin{equation}
\bm{J}= \tau 1+i\alpha\bm{i}+i\beta\bm{j}+i\gamma\bm{k}\equiv\bm{h}.
\end{equation}

We therefore conclude that the  vector state $|h\rangle$, the matrix state $\mathbf{Z}$ and the Jones matrix $\mathbf{J}$ are isomorphic to the same quaternion state, $\bm{h}$. 
\section{Properties of quaternion states}

The algebra of the quaternion states, $\bm{h}$, is free from any matrix (or vector) representation. Hence we will base our formalism on the quaternionic form of the nondepolarizing media states.

\subsection{Multiplication of quaternion states}

We can multiply two (or more) $\bm{h}$ quaternions and that results into another  $\bm{h}$ quaternion:

\begin{equation}\label{quat}
\bm{h}=\bm{h_2}\bm{h_1}=\begin{Bmatrix}(\tau_2\tau_1+\alpha_2\alpha_1+\beta_2\beta_1+\gamma_2\gamma_1)1\\
+i(\tau_2\alpha_1+\alpha_2\tau_1+i\beta_2\gamma_1-i\gamma_2\beta_1)\bm{i}\\
+i(\tau_2\beta_1+\beta_2\tau_1-i\alpha_2\gamma_1+i\gamma_2\alpha_1)\bm{j}\\
+i(\tau_2\gamma_1+\gamma_2\tau_1+i\alpha_2\beta_1-i\beta_2\alpha_1)\bm{k}
\end{Bmatrix}
\end{equation}

It is worth noting that the resultant  $\bm{h}$ quaternion is not a four component column vector, it is a single \emph{hypercomplex number} that corresponds to the covariance vector $|h\rangle$ with the following components:
\begin{equation}
|h\rangle=\begin{pmatrix}\tau_2\tau_1+\alpha_2\alpha_1+\beta_2\beta_1+\gamma_2\gamma_1\\
\tau_2\alpha_1+\alpha_2\tau_1+i\beta_2\gamma_1-i\gamma_2\beta_1\\
\tau_2\beta_1+\beta_2\tau_1-i\alpha_2\gamma_1+i\gamma_2\alpha_1\\
\tau_2\gamma_1+\gamma_2\tau_1+i\alpha_2\beta_1-i\beta_2\alpha_1
\end{pmatrix}
\end{equation}

The $\mathbf{Z}$ matrix serves as a short hand multiplication table for $|h\rangle$ vectors \cite{Chipman}:

\begin{equation}\label{mvmul}
|h\rangle= \mathbf{Z_2}|h_1\rangle
\end{equation}
where $|h\rangle, \mathbf{Z_2}, |h_2\rangle$, correspond respectively to   $\bm{h}$ quaternion,  $\bm{h_2}$ quaternion and  $\bm{h_1}$ quaternion. The quaternion algebra of \eqref{quat} based on the multiplication of two hypercomplex numbers offers a compact and simple alternative to the matrix-matrix multiplication \eqref{mmmu}, and to the matrix-vector multiplication \eqref{mvmul}.

\subsection{Transformatıon of a Stokes quaternion}

A three dimensional vector $\bm{v} (=v_1\bm{i}+v_2\bm{j}+v_3\bm{k})$ can be rotated about an axis by an angle $\theta$ as follows:

\begin{equation}
\bm{v'}=\bm{q}\bm{v}\bm{\bar{q}}
\end{equation}
where $\bm{q}$ is a unit real quaternion, and $\bm{\bar{q}}$ is the Hamilton (quaternion) conjugate of $\bm{q}$:

\begin{equation*}
\bm{q}=q_01+q_1\bm{i}+q_2\bm{j}+q_3\bm{k}\end{equation*}
\begin{equation}
\bm{\bar{q}}=q_01-q_1\bm{i}-q_2\bm{j}-q_3\bm{k}\
\end{equation}
($q_0, q_1, q_2, q_3$ are real numbers such that $\bm{q}\bm{\bar{q}}=\bm{\bar{q}}\bm{q}=q_0^2+ q_1^2+ q_2^2+ q_3^2=1$).

From the formula \eqref{ZSZ} it immediately follows that a very similar transformation (rotation) applies to the Stokes quaternion (see also Liu \emph{et al.}\cite{Liu}):
\begin{equation}\label{quattrans}
\bm{s'}=\bm{h}\bm{s}\bm{h}^{\dagger},
\end{equation}
where $\bm{h}^{\dagger}$ is the the Hermitian conjugate of $\bm{h}$:

\begin{equation}
\bm{h}^{\dagger}=\tau^{*}1 +i\alpha^*\bm{i}+i\beta^*\bm{j}+i\gamma^*\bm{k},
\end{equation}
and $\bm{s}$ is the Stokes quaternion that corresponds to the Stokes vector
$|s\rangle= (s_0,s_1,s_2,s_3)^T$:

\begin{equation}
\bm{s}=s_{0}1 +is_{1}\bm{i}+is_{2}\bm{j}+is_{3}\bm{k},
\end{equation}.

$\bm{s'}$ is the transformed (rotated) Stokes quaternion which corresponds to the transformed Stokes vector $|s'\rangle$:
\begin{equation}
|s'\rangle=\mathbf{M}|s\rangle,
\end{equation}
where $\mathbf{M}$ is a nondepolarizing Mueller matrix. 
The proof is straightforward but tedious. By direct multiplication of three quaternions in \eqref{quattrans} it can be shown that the transformed Stokes quaternion $\bm{s'}$ is isomorphic to the transformed Stokes vector $|s'\rangle$.
\subsection{Rotation of the quaternion state}
The Mueller matrix of an optical element can be rotated by an angle $\theta$ in a plane perpendicular the light propagation direction:
\begin{equation}
\mathbf{M}(\theta)= \mathbf{R}(\theta)\mathbf{M}\mathbf{R}(-\theta),
\end{equation}
where
\begin{equation}
\mathbf{R}(\theta)=
\begin{pmatrix}
1&0&0&0\\
0&\cos(2\theta)&-\sin(2\theta)&0\\
0&\sin(2\theta)&\cos(2\theta)&0\\
0&0&0&1
\end{pmatrix}
\end{equation}

Similarly, the covariance vector of a nondepolarizing Mueller matrix can be rotated as follows:
\begin{equation}
|h(\theta)\rangle=\mathbf{R}(\theta)|h\rangle,
\end{equation}
where $|h(\theta)\rangle$ generates $\mathbf{M}(\theta)$. 

The covariance vector of the matrix $\mathbf{R}(\theta)$ is $|r(\theta)\rangle= (\cos(\theta),0,0,-i\sin(\theta))^T$ with the associated quaternion $\bm{r}$:
\begin{equation}
\bm{r}=\cos(\theta)1+\sin(\theta)\bm{k}
\end{equation}

The quaternion $\bm{r}$ is unitary and it is the rotator for the quaternion $\bm{h}$:
\begin{equation}
\bm{h}(\theta)=\bm{r}\bm{h}\bm{r}^{\dagger},
\end{equation}
where $\bm{r}^{\dagger}= \bm{\bar{r}}= \cos(\theta)1-\sin(\theta)\bm{k}$.

\subsection{Other properties and special cases}
\begin{enumerate}
\item 
The norm of the covariance vector is given by,
\begin{equation}
\langle h|h\rangle=\tau\tau^*+\alpha\alpha^*+\beta\beta^*+\gamma\gamma^*=M_{00}
\end{equation}
In the quaternion language, this norm corresponds to the real part of the quaternion $\bm{h}$ multiplied by its Hermitian conjugate:
\begin{equation} {Re}(\bm{h}\bm{h}^\dagger)=\tau\tau^*+\alpha\alpha^*+\beta\beta^*+\gamma\gamma^*=M_{00}
\end{equation}

\item 
The following property can be used to define successive rotations:

\begin{equation}
(\bm{h}_i\bm{h}_j)^\dagger=\bm{h}_j^\dagger \bm{h}_i^\dagger
\end{equation}
For example, if $\bm{h}_1$ and $\bm{h}_2$ are two quaternions corresponding to two rotations, the transformed Stokes quaternion can be written as:

\begin{equation}
\bm{h}_2(\bm{h}_1\bm{s}\bm{h}_1^\dagger)\bm{h}_2^\dagger=(\bm{h}_2\bm{h}_1)\bm{s}(\bm{h}_1^\dagger\bm{h}_2^\dagger)= (\bm{h}_2\bm{h}_1)\bm{s}(\bm{h}_2\bm{h}_1)^\dagger
\end{equation}
which means that $\bm{h}_2\bm{h}_1
$ is the combined rotator.

\item 
If $\tau$ is real and $\alpha, \beta, \gamma$ are pure imaginary then, 
\begin{equation} \bm{h}\bm{h}^\dagger=\bm{h}^\dagger\bm{h}=\tau\tau^*+\alpha\alpha^*+\beta\beta^*+\gamma\gamma^*=\langle h|h\rangle = M_{00}
\end{equation}
and, if $|h\rangle$ is normalized  to unity then,
\begin{equation} \bm{h}\bm{h}^\dagger=\bm{h}^\dagger\bm{h}=1.
\end{equation}
In this case, $\bm{h^{\dagger}}$ is the inverse of $\bm{h}$, and inverse rotation for Stokes quaternion can be written as,
\begin{equation}
\bm{s}=\bm{h}^\dagger\bm{s'}\bm{h}.
\end{equation}
This case corresponds to unitary $\mathbf{Z}$ and unitary $\mathbf{M}$ \cite{KKO}.

\item 
 In general, inverse rotation is related to the Hamilton conjugate of the quaternion $\bm{h}$ which is defined as

\begin{equation}
\bm{\bar{h}}=\tau 1-i\alpha\bm{i}-i\beta\bm{j}-i\gamma\bm{k}.
\end{equation}

Since 
\begin{equation}
\bm{h}\bm{\bar{h}}=\tau^2-\alpha^2-\beta^2-\gamma^2,
\end{equation}
if $(\tau^2-\alpha^2-\beta^2-\gamma^2)>0$ the inverse of $\bm{h}$ can be defined as follows:

\begin{equation}
\bm{h^{-1}}=\frac{\bm{\bar{h}}}{\tau^2-\alpha^2-\beta^2-\gamma^2}
\end{equation}

Similarly the inverse of the Hermitian conjugate of $\bm{h}$ is 

\begin{equation}
\bm{(h^\dagger)^{-1}}=\frac{(\bm{\overline{h^\dagger}})}{(\tau^*)^2-(\alpha^*)^2-(\beta^*)^2-(\gamma^*)^2}
\end{equation}
where $(\bm{\overline{h^\dagger}})=\bm{h^*}=\tau^{*}1-i\alpha^*\bm{i}-i\beta^*\bm{j}-i\gamma^*\bm{k}$.

\item 
 If $\tau, \alpha, \beta$ and $\gamma$ are real, 
\begin{equation} \bm{h}=\bm{h}^{{\dagger}}.
\end{equation}

In this case, transformation of the Stokes quaternion becomes

\begin{equation}
\bm{s'}=\bm{h}\bm{s}\bm{h}.
\end{equation}

This case corresponds to Hermitian $\mathbf{Z}$ and Hermitian $\mathbf{M}$ \cite{KKO}.
\end{enumerate}
\section{Application to the representation of optical media}

\subsection{Coherent linear superposition of Mueller-Jones states and depolarization}

As we have shown, the Mueller-Jones state of a nondepolarizing optical media can be represented by a covariance vector $|h\rangle$, by a matrix state $\mathbf{Z}$, by a Jones matrix $\mathbf{J}$ or by a quaternion $\bm{h}$.

Any linear combination of quaternions is also a quaternion, and a coherent linear combination of Mueller-Jones states can be written as, 
\begin{equation}
\bm{h}=a\bm{h}_1+b\bm{h}_2+c\bm{h}_3\cdots
\end{equation}
The coefficients $a, b, c, \ldots$ are, in general, complex numbers.

If the process is coherent then the Stokes quaternion is subjected to a rotation by the quaternion state $\bm{h}$ associated with the  combined system. If the process is incoherent we have to consider depolarization effects and in this case, the covariance matrix $\mathbf{H}$ associated with a depolarizing Mueller matrix will be of rank $>$ 1, and depolarizing Mueller matrix can be written as a convex sum of at most four nondepolarizing Mueller matrices\cite{Cloude}:
\begin{equation}\label{convexsum}
\mathbf{M}=w_{1}\mathbf{M_1}+w_2\mathbf{M_2}+w_3\mathbf{M_3}+w_4\mathbf{M_4}
\end{equation}
where $\mathbf{M_1},\mathbf{M_2},\mathbf{M_3}$ and $\mathbf{M_4}$ are nondepolarizing Mueller matrices; $w_1, w_2, w_3$, and $w_4$ are real and positive numbers with the condition,
\begin{equation}
w_1+ w_2+ w_3+w_4=1. 
\end{equation}

Decomposition of a depolarizing Mueller matrix into its nondepolarizing components is not unique. In the spectral (Cloude) decomposition \cite{Cloude}, weights $w_i$ are the eigenvalues of the covariance matrix, $\mathbf{H}$, and the component  matrices $\mathbf{M}_i$ are the nondepolarizing Mueller matrices corresponding to the associated eigenvectors of $\mathbf{H}$.

For an incoherent combination, from the linearity of the convex summation of \eqref{convexsum}, we can immediately write a transformation formula for the Stokes quaternion:

\begin{equation}
\bm{s'}=\sum_{i=1}^{4}w_{i}\bm{h}_i\bm{s}\bm{h}_i^{\dagger}
\end{equation}

 The same depolarization scheme given in \cite{KKPA} applies to the quaternion formulation as well.

\subsection{The nondepolarizing Mueller matrix}
The nondepolarizing Mueller matrix can be recovered by shifting from triple quaternion multiplication \eqref{quattrans} to a matrix-matrix-vector multiplication.

Consider the second product of the quaternion rotation in \eqref{quattrans}. The quaternion product $\bm{s}\bm{h}^{\dagger}$ maps to the following matrix-vector product:

\begin{equation}
\bm{s}\bm{h}^{\dagger}\mapsto \mathbf{S}|h^*\rangle=\begin{pmatrix}
s_0 &s_1 &s_2 &s_3\\
s_1 &s_0 &-is_3 &is_2\\
s_2&is_3&s_0&-is_1\\
s_3&-is_2&is_1&s_0
\end{pmatrix}
\begin{pmatrix}
\tau^*\\\alpha^*\\\beta^*\\\gamma^*
\end{pmatrix},
\end{equation}
where $\mathbf{S}$ is the matrix associated with the Stokes quaternion $\bm{s}$.

It can be shown that the order of multiplication can be reversed by means of the $\mathbf{Z}^*$ matrix:
\begin{equation}
\mathbf{S}|h^*\rangle=\mathbf{Z}^*|s\rangle,
\end{equation}
where $|s\rangle$ is the Stokes vector ($|s\rangle=(s_0,s_1,s_2,s_3)^T$).

Since quaternion $\bm{h}$ is associated with the $\bm{Z}$ matrix, the triple quaternion product maps to the following matrix-matrix-vector product:
\begin{equation}
\bm{s'}=\bm{h}\bm{s}\bm{h}^{\dagger}\mapsto |s'\rangle=\mathbf{Z}\mathbf{Z}^*|s\rangle=(\mathbf{Z}\mathbf{Z}^*)|s\rangle=\mathbf{M}|s\rangle,
\end{equation}
where $\mathbf{M}$ is the nondepolarizing Mueller matrix.
Explicit form of the nondepolarizing Mueller matrix in terms of the parameters 
$\tau, \alpha, \beta$ and $\gamma$ can be found in \cite{KKO}.

\subsection{Exponential and differential form of the quaternion state} 
Any quaternion, $\bm{q}=w1+x\bm{i}+y\bm{j}+z\bm{k}$, can be expressed in an exponential form:
\begin{equation}
\bm{q}=|\bm{q}|(\cos \theta+\bm{\hat{u}}\sin\theta)=|\bm{q}|e^{\bm{\hat{u}}\theta},
\end{equation}
where $|\bm{q}|=\sqrt{\bm{q}\bm{\bar{q}}}$,\: $\cos\theta=w/|\bm{q}|$,\:\:$\bm{\hat{u}}=(x\bm{i}+y\bm{j}+z\bm{k})/\sqrt{x^2+y^2+z^2}$,\:\:$\sin\theta=\sqrt{x^2+y^2+z^2}/|\bm{q|}$.

Similarly the quaternion $\bm{h}=\tau 1+i\alpha\bm{i}+i\beta\bm{j}+i\gamma\bm{k}$ can be written in polar form by using the expressions for $\tau, \alpha, \beta$ and $\gamma$ in terms of the spectroscopic parameters $\eta$ (isotropic phase retardation), $\kappa$ (isotropic amplitude absorption), $CD$ (circular dichroism), $CB$ (circular birefringence), $LD$ (horizontal linear dichroism), $LB$ (horizontal linear birefringence), $LD'$ ($45^{\circ}$ linear dichroism) and $LB'$ ($45^\circ$ linear birefringence):

\begin{align}
\tau&=e^{-\frac{i\chi}{2}}\cos\left(\frac{T}{2}\right)\,
&\alpha&=-e^{-\frac{i\chi}{2}}\frac{iL}{T}\sin\left(\frac{T}{2}\right)\\
\beta&=-e^{-\frac{i\chi}{2}}\frac{iL'}{T}\sin\left(\frac{T}{2}\right)\, &\gamma&=e^{-\frac{i\chi}{2}}\frac{iC}{T}\sin\left(\frac{T}{2}\right)
\end{align}
where $\chi=\eta-i\kappa$, $L=LB-iLD$, $L'=LB'-iLD'$,  $C=CB-iCD$, $T=\sqrt[]{L^2 + L'^2 + C^2}$.

If we choose $\theta=T/2$, then the quaternion $\bm{h}$ can be written as,
\begin{equation}
\bm{h}=e^{\bm{\hbar}},
\end{equation}
where $\bm{\hbar}$ is another quaternion which can be written as
\begin{equation}
\bm{\hbar}=\frac{-iT}{2}(\chi 1+iL\bm{i}+iL'\bm{j}-iC\bm{k}).
\end{equation}

The quaternion $\bm{\hbar}$ corresponds to the differential $\mathbf{z}$ matrix \cite{KKO},

\begin{equation}
\mathbf{z}=\frac{-i}{2}\begin{pmatrix}
\chi&L&L'&-C\\
L&\chi&iC&iL'\\
L'&-iC&\chi&-iL\\
-C&-iL'&iL&\chi
\end{pmatrix}.
\end{equation}

Now we can differentiate $\bm{h}$ with respect to $l$ (distance along the propagation of light)
\begin{equation}
\frac{\mathrm{d}\bm{h}}{\mathrm{d}l}=\frac{\mathrm{d}
\bm{\hbar}}{\textrm{d}l}\bm{h}=\frac{\bm{\hbar}}{l} \bm{h},
\end{equation}

This equation is a reformalism of the Stokes-Mueller differential formalism \cite{ArteagaRev} and it can be compared with the well known quaternion differentiation formula:
\begin{equation}
\bm{\dot{q}}=\frac{1}{2}\bm{\omega}\bm{q}
\end{equation}
where $\bm{\omega}$ is the angular velocity. Hence, $2\bm{\hbar}/l$ can be interpreted as the angular velocity in the rotation by an angle $\theta$ of the quaternion state through the medium.

\section{Conclusion}
It is shown that the Stokes-Mueller formalism can be reformulated in terms of quaternions, and the
quaternion approach is more suitable for the formalism of Mueller-Jones states that we have recently described. The vector state $|h\rangle$, the matrix  state $\mathbf{Z}$ and the Jones matrix $\mathbf{J}$ are consistantly isomorphic to the same quaternion state $\bm{h}$. 

Mueller transformation of the Stokes vector turns out to be a familiar rotation generated by quaternion rotators. Particularly, if $\mathbf{M}$ is a unitary matrix then $\alpha,\beta$ and $\gamma$ are pure imaginary numbers and the quaternion state, $\bm{h}$, becomes a real quaternion. In this case, the quaternion rotation of the Stokes quaternion can be conceived as a three dimensional rotation on the Poincar\'{e} sphere.

In short, quaternion algebra  embraces all views  of the Stokes-Mueller formalism, including coherent linear combination of Mueller-Jones states and depolarization. 

\section{Funding Information and acknowledgement}
Ministerio de Economía y Competitividad
(MINECO) (CTQ2013-47401-C2-1-P, FIS2012-38244-
C02-02). One of the authors (M. Ali Kuntman) thank Dr. Adnan K\"{o}\c{s}\"{u}\c{s}.
\bibliography{sample}

\end{document}